\documentclass[prl,twocolumn,superscriptaddress,groupedaddress,nofootinbib]{revtex4}
\usepackage{color}
\usepackage{graphicx}
\usepackage{dcolumn}
\usepackage{bm}
\usepackage{amssymb}
\usepackage{amsmath}
\usepackage{latexsym}
\usepackage{float}
\usepackage{ifthen}
\usepackage{caption,subfig}
\usepackage{enumerate}
\usepackage{url}
\usepackage{caption,subfig}

\def\ba{\begin{eqnarray}}
\def\ea{\end{eqnarray}}
\def\mpl{M_{\rm Pl}}
\def\e{{\epsilon}}

\def\L{\mathcal{L}}
\def\({\left(}
\def\){\right)}
\def\ie{{\it i.e. }}

\def\nn{\nonumber}
\def\p{\partial}
\def\mn{_{\mu \nu}}

\def\stu{St\"uckelberg }
\def\p{\partial}

\def\<{\langle}
\def\>{\rangle}

\def\d{\mathrm{d}}

\def\gd{g_{\rm eff}}
\def\gu{g^{\rm eff}}

\newcommand{\para}[1]{\par\vspace{2mm}\noindent{\bf\emph{{#1}}}.---}

\begin{document}

\title{Vielbein to the Rescue?}
\author{Claudia de Rham}
\address{CERCA \& Department of Physics, Case Western Reserve University, 10900 Euclid Ave, Cleveland, OH 44106, USA}
\author{Andrew J. Tolley}
\address{CERCA \& Department of Physics, Case Western Reserve University, 10900 Euclid Ave, Cleveland, OH 44106, USA}
\date{\today}

\begin{abstract}
Non--minimal matter couplings have recently been considered in the context of massive gravity and multi--gravity. These couplings are free of the Boulware--Deser ghost in the decoupling limit and can thus be considered within an Effective Field Theory setup. Beyond the decoupling limit the ghost was shown to reemerge in the metric formulation of the theory. Recently it was argued that this pathology is absent when formulated in terms of unconstrained vielbeins. We investigate this possibility and show that the Boulware--Deser ghost is always present beyond the decoupling limit in any dimension larger than two.
We also show that the metric and vielbein formulations have an identical ghost--free decoupling limit. Finally we extend these arguments to more generic multi--gravity theories and argue that for any dimension larger than two a ghost is also present  in the vielbein formulation whenever the symmetric vielbein condition is spoiled and the equivalence with the metric formulation is lost.
\end{abstract}

\maketitle

\para{Introduction}
With the successful development of ghost--free massive gravity \cite{deRham:2010ik,deRham:2010kj}, bi--gravity \cite{Hassan:2011zd} and multi--gravity \cite{Hinterbichler:2012cn} theories (see Refs.~\cite{Hinterbichler:2011tt,deRham:2014zqa} for reviews), the question has naturally arisen of whether there are more general consistent ghost--free couplings to matter. Unlike GR, non--minimal couplings to both metrics are not forbidden by any symmetry, and will inevitably arise in any low energy effective field theory of massive gravity or multi--gravity. A number of possible phenomenological couplings to matter have been explored in the literature \cite{Akrami:2013ffa,Akrami:2014lja,Noller:2014sta,Aoki:2014cla,Enander:2014xga,Schmidt-May:2014xla,Gumrukcuoglu:2014xba,Solomon:2014iwa,Mukohyama:2014rca,Gao:2014xaa,
Blanchet:2015sra,Comelli:2015pua,Gumrukcuoglu:2015nua,Koyama:2015vza} and their theoretical consistency was explored in  \cite{Yamashita:2014fga,deRham:2014naa,Hassan:2014gta,deRham:2014fha,Soloviev:2014eea,Heisenberg:2014rka,Hinterbichler:2015yaa,Matas:2015qxa}.

As is well known, all multi--gravity theories have a low strong coupling scale $\Lambda$ which lies parametrically between the graviton mass and the Planck mass. For massive gravity theories with a fixed reference metric, in $D$ dimensions the scale is $\Lambda = (m^2 \mpl^{(D-2)/2} )^{2/(D+2)}$. For multi--gravity it is a generalization which is typically dominated by the lowest non--zero graviton mass and the lowest Planck scale  \cite{deRham:2013awa,Noller:2015eda}.
If these theories are viewed in terms of an effective field theory (EFT), then $\Lambda$ may be taken to be the cutoff of the EFT. In which case, we only require for consistency of the theory, the absence of ghosts below the cutoff scale $\Lambda$. In particular any non--minimal matter coupling which remains ghost--free in the decoupling limit (DL) $\mpl \rightarrow \infty$ keeping $\Lambda$ fixed is considered acceptable from an EFT point of view. In a recent work precisely such a coupling was proposed in the metric formulation of massive gravity and bi--metric theories, and amounts to an effective `composite' metric $\gd$ in  \cite{deRham:2014naa},
\ba
\label{eq:effMetric}
\gu\mn= g\mn +2 \alpha  g_{\mu \alpha}\(\sqrt{g^{-1}f}\)^\alpha_{\ \nu}+\alpha^2 f\mn\,,
\ea
where $g\mn$ and $f\mn$ are two metrics (in bi--gravity they are the two dynamical metrics, while in massive gravity $g\mn$ can represent the dynamical metric and $f\mn$ the Minkowski reference metric).

In terms of the effective metric \eqref{eq:effMetric}, the coupling to matter then takes the standard form,
\ba
\mathcal{L}_{\rm matter}&=& \mathcal{L}_g\(g, \p \psi_g, \psi_g\)+\mathcal{L}_f\(f, \p \psi_f, \psi_f\)\nn \\
&+&\mathcal{L}_{\rm eff}\(\gd, \p \chi, \chi\)\,,
\ea
where $\psi_g$, $\psi_f$ and $\chi$ symbolize different matter sectors.
The generalization of the coupling to the effective metric \eqref{eq:effMetric} was also considered for multi--gravity in  \cite{Noller:2014sta}. In the vielbein formalism\footnote{See Refs.~\cite{Nibbelink:2006sz,Chamseddine:2011mu,Hinterbichler:2012cn} for the formulation of massive gravity in the vielbein.} the effective vielbein takes the remarkable simple form
\ba
e_{\rm eff} = \sum_{I=1}^N \alpha_{I} e^{(I)}\,,
\ea
where $e^{(I)}$ is the vielbein for the metric $g\mn^{(I)}$, and the $\alpha_{I}$ are arbitrary dimensionless coefficients.

\para{Vainshtein}
As we have emphasized, from an EFT point of view the absence of ghosts below $\Lambda$ is a sufficient criterion. Nevertheless, it has been traditional to ask for a tighter restriction in massive gravity theories. If the additional interactions were ghost--free nonlinearly, then the Vainshtein mechanism allows us to make sense of the theory above the scale $\Lambda$, by redressing the cutoff scale around certain backgrounds. Thus in order to address the viability of the Vainshtein mechanism, it is natural to ask whether there exist any non--minimal couplings to matter which are ghost--free nonlinearly. In the case of the composite metric proposal of \cite{deRham:2014naa} this was definitively shown to be ghostly \cite{deRham:2014fha,Soloviev:2014eea}. More recently, it has been argued that the unconstrained vielbein version of this interaction was ghost--free~\cite{Hinterbichler:2015yaa}. A priori this is not a contradiction since the vielbein and metric formulations are not equivalent. This is because integrating out Lorentz transformations that encode the extra variables in the vielbein no longer imposes the symmetric vielbein (or Deser--van Nieuwenhuizen) condition \cite{Deser:1974cy,Hoek:1982za}, but a non--trivial condition which depends on the matter content. As such the metric formulation of the unconstrained vielbein formulation would contain highly non--trivial non--minimal matter couplings and would differ from a effective coupling to \eqref{eq:effMetric}.

\para{Boulware--Deser ghost}
Before moving to the core of the argument, we first clarify what we mean by the Boulware--Deser ghost as first discovered in \cite{Boulware:1973my}. The Boulware--Deser (BD) ghost is usually associated with the loss of the primary constraint generated by the lapse $N$. However  one can always push the problem to the level of the secondary constraints by introducing auxiliary variables. For instance one may consider the Hamiltonian $H=N R_0+N^2 Q$, where $R_0$ and $Q$ are functions of the dynamical phase space variables. In this formulation it is clear that $N$ does not generate a primary constraint. For this example, we can restore that primary constraint at the price of spoiling the secondary constraints by introducing an auxiliary variable $\sigma$ and conjugate momentum $p_\sigma$, with $H=N(R_0+\sigma Q)-\frac 14 \sigma^2 Q + \lambda p_\sigma$, where $\lambda$ is a Lagrange multiplier that ensures that $p_\sigma=0$. In that formulation, the primary constraint associated with the lapse $N$ survives $C_{1,I}=R_0+\sigma Q \approx 0$. In addition, there we have another primary constraint namely $C_{2,I}=p_\sigma\approx 0$. The problem manifests itself at the level of the secondary constraints, where the one associated with $C_{2,I}$ now involves the lapse:
\ba
C_{2,II}=\{C_{2,I},H\}=-\frac{\p H}{\p \sigma}= Q \(N -\frac 12 \sigma\)\approx0\,,
\ea
and this equation can be solved for the lapse $N$ rather than one of the dynamical variables and so is no longer a true constraint. Similarly the secondary constraint associated with $C_{1,I}$ can now be solved for the Lagrange multiplier $\lambda$ and we thus loose two secondary constraints.  \\

In the formulation of \cite{Hinterbichler:2015yaa} the primary constraint associated with $N$ exists in all dimensions, but the existence of secondary constraints needed to ensure the correct number of degrees of freedom in $D>2$ dimensions are not guaranteed\footnote{We thank K.~Hinterbichler and R.~Rosen for discussions on this point.}. In that case the analogue of the auxiliary variable $\sigma$ are the Lorentz rotations (to be defined below) and the number of secondary constraints associated with them increases with the number of dimensions. In this language we may thus wonder if the number of ghost would increase with the number of dimensions. The answer is no. The reason for that is while all the $\frac 12 (D-2)(D-1)$ secondary constraints associated with the rotations involve the lapse, it is always possible to reformulate them in a way where only one of these equations depend on the lapse and the others are free of the lapse and therefore genuinely generate $\frac 12 (D-2)(D-1)-1$ secondary constraints. The second secondary constraint which is lost is the same one as that associated with $C_{1,I}$ which can be solved for one the Lagrange multipliers $\lambda$ and so a total of two secondary constraints are lost.
So even though the number of auxiliary variables increases with the number of dimensions, the number of ghosts does not. In this sense the ghost found here can be attributed to the BD ghost as opposed to the ghosts found in \cite{deRham:2014tga} and in \cite{deRham:2015rxa} which do increase with the number of dimensions. \\

In this manuscript we shall diagnoze the problem directly at the level of the primary constraint by integrating out the auxiliary variables (namely the rotations). In this language it is more manifest that we are still dealing with the BD ghost, i.e. a failure of the Hamiltonian to be linear in $N$.

\para{Counting}
To understand the loss of the BD constraint, let us first consider the massive gravity limit. Since massive gravity arises as a consistent decoupling limit of multi--gravity theories, the existence of a ghost in the massive gravity limit implies a ghost in any multi--gravity extension by the usual reasoning.

In what follows we use the notation that greek indices represent space--time indices $\mu, \nu=0,\cdots,D-1$, $i,j=1,\cdots, D-1$ run over space dimensions, $A,B,\cdots=0, \cdots ,D-1$ run over all the Lorentz directions while $a,b,\cdots=1, \cdots ,D-1$ run over only the space Lorentz directions.

We begin with the vielbein formulation of massive gravity  with the usual EH kinetic term and with matter coupling to the composite vielbein $e+ \alpha f$.
Working in unitary gauge $f_{\mu}^{\ A} = \delta_{\mu}^{\ A}$, we may decompose the dynamical vielbein $e^a$ into a Lorentz boost/rotation of an upper--triangular vielbein which contains the same number of variables as the metric \cite{Hinterbichler:2012cn}
\ba
\label{eq:TriangleVielbein}
e^{\ A}_{\mu} = \Lambda^A{}_B E_\mu^{\ B}
\ea
where $E^A$ is upper--triangular along the time--direction and symmetric in the space--space directions.
Specifically it takes the form
\ba
&& E^0 = N \d t \\
&& E^a = N^i  E_i^{\ a} \d t + E_i^{\ a} \d x^i\,,
\ea
where $E_{i}^{\ a}=E_{\ a}^{\ i}$. The spatial 3--metric is $g_{ij}=E_i^{\ a}E_j^{\ b} \delta_{ab}$. In general we may factorize the Lorentz transformations into
\ba
\Lambda^A{}_B= \Lambda^A{}_C(\text{boost}) \Lambda^C{}_B(\text{rotations}) \, ,
\ea
where the rotations act only in the space--space directions.

For instance in $D=3$ dimensions, we express the nine components of the vielbein in terms of a lapse $N$, two shift $N^i$, two boosts $v^a$, one rotation $r$ and the three components of a  symmetric spatial vielbein $E_i^{\ a}= E_a^{\ i}$
\ba
e^{\ A}_\mu=\(
\begin{array}{cc}
N \gamma +N^i e_i^{\ a} v_a & N v^a+N^i e_i^{\ b}\(\delta_b^a+\frac{1}{\gamma+1}v_b v^a\) \\
e_i^{\ a}v_a & e_i^{\ b}\(\delta_b^a+\frac{1}{\gamma+1}v_b v^a\)
\end{array}\)\,,\nn
\ea
where $\gamma$ is given by $\gamma=\sqrt{1+v^2}$ and $e_i^{\ a}$ is the rotated version of $E_i^{\ a}$,
\ba
\label{Lambdar3d}
e_i^{\ a}=\(\Lambda_r\)^a_{\ b}E_i^{\ b}
=\(\begin{array}{cc}
\gamma_r & r\\
-r & \gamma_r
\end{array}\)\(\begin{array}{cc}
E_1^{\ 1} & E_1^{\ 2} \\
E_1^{\ 2} & E_2^{\ 2}
\end{array}\)\,,
\ea
with $\gamma_r=\sqrt{1-r^2}$.
In arbitrary dimensions, this decomposition may be summarized as
\ba
 D^2 \text{ (vielbein)} &=& (D-1) \text{ (boosts)}  \\
&+&\frac{1}{2} (D-2)(D-1)\text{ (rotations)}  \nn\\
&+& D\,  \text{(lapse and shift)}\nn\\
&+& \frac{1}{2}D(D-1) \text{ (spatial metric)}\,, \nn
\ea
where the spatial metric is given uniquely in terms of the spatial symmetric vielbein $E_i^{\ a}$.\\

Now the crucial property is that as long as we choose the EH kinetic term, the $(D-1)$ boosts and the $\frac{1}{2}(D-2)(D-1)$  rotations drop out of the kinetic term, meaning that in unitary gauge they are non--dynamical variables which do not enter the phase space symplectic form. In addition, the lapse and shift $N$, $N^i$ are also non--dynamical, leading to a total of $\frac{1}{2} D(D+1)$ auxiliary variables. The $\frac{1}{2}D(D-1)$ $E_{i}^{\ a}$ are dynamical and after integrating out the spin--connection all receive a kinetic term. If none of the auxiliary variable played the role of a Lagrange multiplier,
 the total dimension of the phase space would be
\ba
\text{Phase space dimension} = 2 \times \( \frac{1}{2} D(D-1)\) \, .
\ea
On the other hand the number of degrees of freedom of a massive graviton in $D$ dimensions, is that of a massless one in $D+1$ dimensions, namely $\frac{1}{2}D(D-1) -1$. As a result we would obtain in any dimension one additional degree of freedom which is the BD ghost. We thus require for the absence of the BD ghost that one combination of the auxiliary variables acts as a Lagrange multiplier.  \\

In \cite{Hinterbichler:2015yaa} the existence of this additional constraint was argued as follows. Since the Hamiltonian is clearly linear in the shift $N^i$, we may use the shift constraint to solve for the $D-1$ Lorentz boosts in a manner which preserves linearity in $N$. This then ensures that $N$ imposes a constraint, and so naively the BD ghost must be absent. This argument is correct in $D=2$ since their are no rotations. However in all dimensions $D>2$, it is necessary to further check that it is possible to integrate out the rotations in a manner which preserves the linearity in $N$. Equivalently we may use the fact that since the rotations do not have a kinetic term, we can introduce a momentum conjugate for them and an additional primary constraint that they have vanishing momenta. For the counting to work out, these must lead to secondary constraints. The secondary constraints associated with the vanishing of the momentum conjugate for the rotations are none other than the space-space part of the symmetric vielbein condition and the requirement that these are truly constraints is equivalent to asking that the rotations can be integrated out in a manner which preserves linearity in $N$. The arguments of \cite{Hinterbichler:2015yaa} assumed the secondary constraints which fix the rotations would arise and this is what we find not to be the case.

\para{Symmetric Vielbein Condition}
In the case of massive gravity with minimal coupling to matter the secondary constraint exists and this can be understood as follows: As is well known, in the case of minimal coupling to matter, varying the action with respect to the boosts and rotations reproduces the symmetric vielbein condition\footnote{There generically exist other branches of solutions of these equations in which the symmetric vielbein condition does not hold (see for example \cite{Deffayet:2012zc,Banados:2013fda}), however these are disconnected branches in which ghosts are present.}
\ba
e_{\mu}^{\ A} f_{\nu}^{\ B} \eta_{AB} = e_{\nu}^{\ A} f_{\mu}^{\ B} \eta_{AB}  \, .
\ea
If we consider only the space--space part of this equation, $e_{i}^{\ A} f_{j}^{\ B} \eta_{AB} = e_{j}^{\ A} f_{i}^{\ B} \eta_{AB}$ then those constitute $\frac{1}{2}(D-2)(D-1)$ equations. It is precisely these space-space parts that arise as the secondary constraints associated with the vanishing of the momentum conjugate to the rotations. \\

Crucially since $e_i^{\ A}$ is independent of the lapse and shift, then these $\frac{1}{2}(D-2)(D-1)$ equations may be solved algebraically to determine the $\frac{1}{2}(D-2)(D-1)$ Lorentz rotations in terms of $E_i^{\ a}$ and the boosts $v^a$, and these solutions are independent of $N$ and $N^i$. In other words these are truly constraints on the system which can be used to fix the rotations. Substituting back into the action, the Hamiltonian remains linear in $N^i$ and the associated $N^i$ constraints may be used to solve for $v^a$. This leaves behind the $N$ which imposes the final primary constraint which removes the BD ghost. \\

In the case of the non--minimal matter coupling, the symmetric vielbein condition is lost and gets sourced instead by the stress--energy tensor of the non--minimally coupled matter fields. Since this stress--energy depends on the shift and lapse, it follows that these $\frac{1}{2}(D-2)(D-1)$ equations can be solved to determine the $\frac{1}{2}(D-2)(D-1)$ Lorentz rotations in terms of not only $E_i^{\ a}$ and the boosts $v^a$, but also $N$ and $N^i$. This dependence on the Lagrange multipliers means that these are not true constraints any more (or at least not all of these equations are constraints). Plugging these expressions back in the Hamiltonian, the latter will no longer be linear in the shift and the lapse, hence spoiling the usual constraint that removes the BD ghost. This argument works in all dimensions $D>2$ and for arbitrary matter fields. For concreteness, we focus in $D=3$ dimensions, with the understanding that adding dimensions to the theory will not alleviate the problem (quite the opposite).

\para{Integrating out the auxiliary variables in massive gravity}
In order to better compare with the case of the non--minimal coupling, we start by looking at the mass term explicitly.
To simplify the discussion, we consider the following mass term
\ba
H_{\rm mass}=\mpl \frac {m^2}2 \epsilon_{ABC} \ e^A\wedge f^B \wedge f^C \,.
\ea
Expressed in terms of the dynamical and auxiliary variables, the Hamiltonian for this mass term takes the form
\ba
&&H_{\rm mGR} = N^0 R+N^i R_i\nn \\
&&+\mpl m^2\left[ N \gamma
+e_i^{\ a} \(\delta^i_a+\frac{v^iv_a}{1+\gamma}\)
+N^i e_i^{\ a}v_a \right]\,.
\label{Hmass}
\ea
We can now integrate out the auxiliary variables as follows: We first solve the equation of motion for the shift in terms of the boost $v^i$ which will then be expressed in terms of the dynamical variables $E_i^{\ a}$ and the rotations,
\ba
\mpl  m^2 e_{i}^{\ a}v_a=-R_i\,.
\ea
Even though $v_j$ depends on the rotations, the rotation matrix $\Lambda_r$ defined in \eqref{Lambdar3d} drops out of the inner product $v^2$. As a result the Lorentz factor $\gamma$ is independent of the rotation which only enter in $e_i^{\ a} \(\delta^i_a+\frac{v^iv_a}{1+\gamma}\)$. Since the lapse does not enter that expression, we can easily integrate out the rotation without involving the lapse. As a result the Hamiltonian remains linear in the lapse after integration of the auxiliary variables.

\para{Non--minimal coupling}
This result only holds for a minimal matter coupling which does not include the Lorentz \stu fields. When matter is chosen to couple to the composite vielbein, $e_{\rm eff}=e+ \alpha f$, this coupling breaks local Lorentz invariance, then the matter Lagrangian contains explicit dependence on the Lorentz boosts and rotations. Varying the action with respect to these Lorentz boosts/rotations gives now a modification of the symmetric vielbein condition which includes a contribution from the matter Lagrangian. However generically this condition depends explicitly on $N$ and $N^i$. We may continue to solve the space--space component of this equation for the Lorentz rotations, but the solution will now depend on $N$. Substituting back in the action, this will generate non--linearities in $N$ not present for minimal couplings. These destroy the existence of the BD constraint as we shall see explicitly below. Rather than re--deriving the modification of the symmetric vielbein condition, we follow the same procedure as previously and first solve the shift equation for the boosts and then integrate out the rotation.\\

For definiteness we consider a scalar field $\chi$. Its potential goes as $\sqrt{-g_{\rm eff}}\, V(\chi)$ and can always be absorbed in a redefinition of the mass terms and is thus ignored here. Moreover to go straight to the point, we focus on the ultra--local limit. This is justified since the gradient terms for matter cannot possibly compensate the contributions from the kinetic term. In other words, if a constraint is lost in the ultra--local limit then it is lost in the whole theory.
Then in terms of the momentum conjugate $p_\chi$ associated with the scalar field $\chi$, the contribution to the Hamiltonian from the non--minimal matter coupling is
\ba
\label{eq:Hmatter1}
H_{\rm matter}=\frac12 \frac{N_{\rm eff}}{\sqrt{\det g^{\rm eff}_{ij}} }p_\chi^2\,,
\ea
where $N_{\rm eff}$ is the effective lapse of the composite vielbein  $g^{\rm eff}_{ij}$ is the effective spatial metric. These effective quantities can be expressed in terms of the shift, lapse, boost, rotation and spatial symmetric vielbein of the dynamical vielbein (see for instance Eq.~(7) of Ref.~\cite{Hinterbichler:2015yaa}). As mentioned in Ref.~\cite{Hinterbichler:2015yaa}, the effective lapse $N_{\rm eff}$ is linear in the shift and lapse $N^i$ and $N$ and $g^{\rm eff}_{ij}$ is independent of the shift and lapse. Notice however that both $N_{\rm eff}$ and $g^{\rm eff}_{ij}$ depend non--trivially on the rotations.

The exact expression of the effective quantities in terms of the dynamical metric is complicated but
for the purpose of this discussion, it is sufficient to perform an expansion in $\alpha$ (recalling that $\alpha=0$ corresponds to the minimal matter coupling). Once again for concreteness we also work in $D=3$ dimensions, although the results are generalizable to any $D>2$ dimensions.

In terms of the lapse, shift, boost and spatial vielbein, the non--minimal matter Hamiltonian \eqref{eq:Hmatter1} takes the form to second order in the parameter $\alpha$
\ba
H_{\rm matter}&=&\frac 12 \frac{N}{\sqrt{g}} p_\chi^2 + \frac{\alpha\, p_\chi^2}{ \sqrt{g} }\Big[(N^i v_i+\gamma)
-\frac{N}{2\sqrt{g}}\, G(e,v^i)
\Big]\nn  \\
&+&\frac{\alpha^2 p_\chi^2}{2g}\Bigg[(1-2\gamma^2)e_i^{\ i}+(1+2 \gamma)e_{i}^{\ a}\frac{v^i v_a}{1+\gamma}\nn \\
&+&N^i v^j e_{k\ell}\(\delta^k_i\delta^\ell_j-2\gamma \delta_{ij}\delta^{k\ell}+2\delta_{ij}\frac{v^k v^\ell}{1+\gamma}\)\nn\\
&+&N Q(e,v^i)
\Bigg]
+\mathcal{O}\(\alpha^3\)\,,
\ea
where $g$ denotes the determinant of the spatial metric, $g=\det g_{ij}$ and is invariant under rotations. In $D=3$ dimensions, $g=(\e^{ij}\e_{ab}e_i^{\ a}e_j^{\ b})^2=(\e^{ij}\e_{ab}E_i^{\ a}E_j^{\ b})^2$. $G$ and $Q$ are functions of the boost and the rotation given by
\ba
\label{eq:G}
G=\e^{ij}\e_{ab}\(\delta_{i}^{a}+\frac{v_iv^a}{1+\gamma} \)e_{j}^{\ b}\,,
\ea
and
\ba
Q&=&e_{ij}e_{k\ell}\Bigg[
-(1+\gamma)(8-3 \gamma-6\gamma^2+2 \gamma^3)\delta^{ij}\delta ^{k\ell}\nn\\
&+&(12+7\gamma-4\gamma^2)\delta^{ij}v^kv^\ell+(1+\gamma)(-3+\gamma)\delta^{ik}v^j v^\ell\nn\\
&-&(1+\gamma)\delta^{i\ell}\delta^{kj}-(2+\gamma)\delta^{i\ell}v^kv^j-v^iv^jv^kv^\ell
\Bigg]\,.
\ea

\para{Integrating out the auxiliary variables}
We may now integrate out the auxiliary variables as we did in the case of massive gravity. The shift equation leads to the equation for the boosts:
\ba
&& \mpl m^2 e_{i}^{\ a}v_a+\alpha\frac{p_\chi^2}{\sqrt g}v_i\\
&&+\alpha^2 \frac{p_\chi^2}{2g}v^j e_{k\ell}\(\delta^k_i\delta^\ell_j-2\gamma \delta_{ij}\delta^{k\ell}+2\delta_{ij}\frac{v^k v^\ell}{1+\gamma}\)=-R_i\,,\nn
\ea
which can be easily solved perturbatively in $\alpha$ to obtain an expression for the boosts in terms of the symmetric vielbein $E_{ij}$ and the rotation. Next we derive the equation of motion for the rotation. To simplify this derivation, we perform a perturbation in $R_i$. To quadratic order in $R_i$, the rotation $r$ defined in \eqref{Lambdar3d} satisfies
\ba
r&=&\frac{\epsilon^{ij}R_i E_{j}^{\ k}R_k}{8\mpl^2m^4[E]\sqrt{g}}\Bigg[1+\frac{\alpha p_\chi^2}{4 \mpl m^2 g}\(4N-[E]\)\\
&-&\frac{\alpha^2 p_\chi^2}{2\mpl m^2 g^2}\Bigg\{
\sqrt{g}\(\frac 12 [E^2]-\frac 32 [E]^2+6N[E]\)\nn\\
&-&\frac{p_\chi^2}{8\mpl m^2}\(\frac12[E^2]+\frac 12 [E]^2-8 [E]N+4N^2\)
\Bigg\}
\Bigg] \nn \\
&+&\mathcal{O}\(R_i^3/\mpl^3m^{6}, \alpha^3\)\,,\nn
\ea
where we used the notation $[E]=E_i^{\ i}$ and $[E^2]=E_i^{\ j}E_j^{\ i}$.
We now see explicitly that the rotation depends on the lapse $N$ and so integrating out the rotation spoils the linearity of the Hamiltonian in the lapse. To leading order in $\alpha$, the rotation is `only' linear in the lapse but to second order in $\alpha$ the non--linear dependence in the lapse becomes fatal. Integrating out the rotation, we find the Hamiltonian now carries a piece non--linear in the lapse\footnote{Even though we have only solved the rotation to quadratic order in $R_i$ it is sufficient to derive the Hamiltonian to quartic order in $R_i$ because the Hamiltonian does not carry any piece in the rotation which is not also multiplied by $R_i^2$. In other words, if we set $R_i \to \epsilon R_i$ and the rotation $r\to \epsilon^2 r$, then the Hamiltonian goes symbolically as $H=1+ \epsilon^2 R_i^2 + \epsilon^4 (R_i^4+ r R_i^2)$, so we only need to solve the rotation to quadratic order in $R_i$ to know the Hamiltonian to quartic order in $R_i$.},
\ba
H_{\rm total}&=&H_0+H_1 N\\
&+&\frac{\alpha^2 p_\chi^4}{64 \mpl^5 m^{10}[E]g^3}\(\epsilon^{ij}R_i E_{j}^{\ k}R_k\)^2 N^2 \nn \\
&+& \mathcal{O}\(R_i^5/\mpl^5 m^{10}, \alpha^3\)\,.\nn
\ea
As a result the Hamiltonian is actually genuinely non--linear in the lapse as soon as the non--minimal coupling is considered and the BD constraint is lost in the full theory.

\para{Extension to bi--gravity and multi--gravity}
These arguments may easily be extended to multi--gravity theories. For instance in the case of bi--gravity, we must add to the set of auxiliary variables, the lapse and shift of the $f$ metric, $M, M^i$. The full set of auxiliary variables are then the $\frac{1}{2} D(D+3)$ variables $\{ \mu^I \}= \{N^{\mu},M^{\mu},\Lambda^{\ a}_b \}$. Since bi--gravity caries one copy of diffeomorphisms, we are guaranteed the existence of $D$ first class constraints. This means that to ensure the absence of a BD ghost, we require that the Hamiltonian remains linear in the two lapses $N^0$ and $M^0$ and the shifts $M^i$ after integrating out all the Lorentz \stu $\Lambda^{\ a}_b$. That this is not the case in the presence of matter coupled to the composite vielbein follows either by direct calculation or by recognizing that since massive gravity arises in the decoupling limit $M_f \rightarrow 0$, and in this limit $M^{\mu}$ decouples.
Since this is a smooth limit, if the Hamiltonian was linear in the lapse $N$ in bi--gravity then it would be so in the massive gravity limit which is a contradiction. In other words a BD ghost cannot reappear from a consistent decoupling limit. The same argument rules out all the multi--gravity extensions.

\para{Decoupling Limit}
Although as we have emphasized, the unconstrained vielbein formulation is fundamentally different than the metric formulation, both have the same $\Lambda$ decoupling limit. The vielbein derivation of the decoupling limit including the vector degrees of freedom was given for massive gravity in \cite{Ondo:2013wka,Gabadadze:2013ria}, bi--gravity in \cite{Fasiello:2013woa} and more recently in multi--gravity in \cite{Noller:2015eda}.
To take this limit, we move out of unitary gauge and work with the reference vielbein in the form
\ba
f^a = (e^{\lambda})^a{}_b \d \phi^b\,,
\ea
where $\lambda_{ab}$ are the Lorentz \stu fields and $\phi^a$ the diffeomorphism \stu fields.
We may for convenience choose $e^a$ to respect the symmetric vielbein condition. We are allowed to do this because on introducing the Lorentz \stu fields, we recover a local Lorentz symmetry which may then be used to set this condition on the vielbeins. With this condition, $e^a$ then encode the usual variables of the metric.
Then the scalings which determine the decoupling limit, which as usual are determined by canonically normalizing the quadratic fluctuations, are in general dimensions,
\ba
&&  e^a = 1^a + \frac{1}{\mpl^{(D-2)/2}}v^a \, ,  \\
&& \phi^a = x^a - \frac{1}{m \mpl^{(D-2)/2}} A^a - \frac{1}{\Lambda^{2+(D-2)/2}} \partial^a \pi  \, , \\
&& \lambda_{ab} = \frac{1}{m \mpl^{(D-2)/2}} \hat \lambda_{ab}  \, ,
\ea
were $\Lambda^{(2+(D-2)/2)}= m^2 \mpl^{(D-2)/2}$.
In particular, in this limit $\mpl \rightarrow \infty$ for fixed $\alpha$, the composite vielbein is given by
\ba
e^a_{\mu}+\alpha f^a_{\mu} = ((1+\alpha)\delta^a_{\mu}-\alpha  \Pi^a{}_{\mu})
+  {\cal O}\left( \frac{1}{m \mpl^{(D-2)/2}} \right) \, ,
\ea
where $\Pi_{\mu\nu} = \Lambda^{-(2+(D-2)/2)} \partial_{\mu}\partial_{\nu} \pi$.
Following \cite{deRham:2014naa} we assume that the matter which couples to this effective metric, represented by $\chi$, does not scale. In which case, in the limit $\mpl \rightarrow \infty$, for fixed $\Lambda$, the matter Lagrangian remains finite with the fields $\chi$ minimally coupled to the composite metric
\ba
g_{\mu\nu}^{\rm eff} = ((1+\alpha) \eta_{\mu \nu} - \alpha \Pi_{\mu\nu})^2\,,
\ea
and crucially $A^a$ and $\lambda_{ab}$ drop out of the matter Lagrangian.
Hence the Lorentz \stu fields enter into the DL precisely in the same way as they do in the case of minimal coupling as given in  \cite{Ondo:2013wka,Gabadadze:2013ria}, which means that they continue to impose the DL version of the symmetric vielbein condition, \ie the equation that determines $\lambda_{ab}$ in terms of $A^a$. The Lorentz \stu fields may then be integrated out to generate the standard vector interactions \cite{Ondo:2013wka,Gabadadze:2013ria}.
The decoupling limit of the composite vielbein formulation is then identical the DL of the composite metric formulation \cite{deRham:2014naa}, and the latter was shown to be ghost--free in \cite{deRham:2014naa}. This ensures that the ghost is absent below and around the scale $\Lambda$. Furthermore it is consistent to take Vainshtein screened solutions with $\Pi \sim 1$ without the ghost being a problem.

\para{Breaking the symmetric vielbein condition in multi--gravity}
Another interesting situation where the symmetric vielbein condition is broken and the equivalence between the metric and vielbein formulation is spoiled arises in the context of multi--gravity \cite{Hinterbichler:2012cn,Scargill:2014wya} and specifically in the presence of `cycles of interactions'. The metric formulation of such theories was conjectured to exhibit a ghost in \cite{Hinterbichler:2012cn} and this was proven directly in the decoupling limit in \cite{Scargill:2014wya}. The vielbein formulation on the other hand breaks the symmetric vielbein condition. As such the equivalence with the metric is broken, leaving open the possibility that the vielbein formulation could be better behaved. To better understand the implications of breaking the symmetric vielbein condition, we consider a specific example in $D=3$ dimensions (see also \cite{Afshar:2014dta}),
\ba
\L&=&M_1 \L_{\rm EH}[e_1]+M_2 \L_{\rm EH}[e_2] \\
& &+ \mpl m^2 \epsilon_{ABC}\Big(c_1 e_1^A\wedge e_1^B \wedge e_2^C\nn \\
&&+c_2 e_2^A\wedge e_2^B \wedge f^C
+c_3 f^A\wedge f^B \wedge e_1^C\Big)\nn\,,
\ea
where for simplicity we consider the ``massive gravity" limit of tri--gravity where one of the vielbein acts as a reference metric $f^A$.
We are dealing with a loop as long as none of the coefficients $c_i$ vanish. If $c_1 c_2 c_3=0$ we are then dealing with a line of interaction rather than a cycle.

Proceeding as before, we can work in unitary gauge for the reference vielbein $f^A_\mu=\delta^A_\mu$ and decompose each of the two dynamical vielbein $e_1$ and $e_2$ into a Lorentz transformation of an upper--triangular vielbein as in Eqns.~(\ref{eq:TriangleVielbein}--\ref{Lambdar3d}). We can now proceed as earlier:  one can use one set of shift equations to solve for one set of boosts. Then the remaining auxiliary variables, namely the  remaining set of boosts and the two rotations can be integrated. \\

In principle these integrations are quite involved but to go straight to the point we can perform a perturbation in the momentum  constraints $R_{i}$. Then all the four boosts start at first order in $R_i$. To zeroth order in $R_i$, the Hamiltonian only includes the rotations and the lapses as auxiliary variables,
\ba
H&=&N_1 R_{1,0}+N_2 R_{2,0}+2m^2\mpl \Big\{c_1 [e_1]+c_3\sqrt{g_2}\quad \\
&+&N_1 (c_1+c_2 \epsilon^{ij}\epsilon_{ab}e_{1,i}^{\ \ a}e_{2,j}^{\ \ b})
+N_2 \(c_3 [e_2]+c_2 \sqrt{g_1}\)
\Big\}\nn\\
&&+\mathcal{O}(R_i)\,,\nn
\ea
where we use the same notation as before, $[e_\sigma]=e_{\sigma,i}^{\ \ i}$, and $\sqrt{g_\sigma}=\epsilon^{ij}\epsilon_{ab}e_{\sigma,i}^{\ \ a}e_{\sigma,j}^{\ \ b}$, and $N_{\sigma=1,2}$ are the two lapses (not to be confused with the shifts $N^i_\sigma$).

Since the boosts vanish to that order in $R_i$, the three symmetric  vielbein conditions would read
\ba
\label{eq:SV}
\mathcal{A}=E^{\ \ a}_{1,[1}E^{\ \ b}_{1,2]}\delta_{ab}=0 \quad{\rm and}\quad r_1=r_2=0\,.
\ea
On the other hand, the equations of motion with respect to the rotation then leads to the modified symmetric vielbein condition
\ba
\label{rot1}
&-&c_2 N_1\gamma_{r_1}\(-\gamma_{r_2}\mathcal{A} +r_2 \sqrt{G_{12}}\)\\
&+&r_1 \(c_1[E_1]+c_2 N_1 (\gamma_{r_2}\sqrt{G_{12}}+r_2\mathcal A)\)\nn=0\\
\label{rot2}
&-&c_2 N_1\gamma_{r_2}\(\gamma_{r_1}\mathcal{A} +r_1 \sqrt{G_{12}}\)\\
&+&r_2 \(c_3 N_2 [E_2]+c_2 N_1 (\gamma_{r_1}\sqrt{G_{12}}-r_1\mathcal A)\)\nn=0\,,
\ea
with $\sqrt{G_{12}}=\epsilon^{ij}\epsilon_{ab}E_{\sigma,i}^{\ \ a}E_{\sigma,j}^{\ \ b}$.
To that order, the symmetric vielbein condition \eqref{eq:SV} is still a solution of the equation, but it is not the most general one. More generally the two equations    \eqref{rot1} and \eqref{rot2} can be solved for $r_1$ and $r_2$ without specifying $\mathcal{A}$.
We can see immediately that if $c_2=0$, then the solution is simply $r_1=r_2=0$ and we recover at least part of the symmetric vielbein condition. If $c_1=0$ or $c_3=0$, one of the rotation still vanishes while the other one is uniquely determined in terms of the $E_{\sigma, i}^{\ \ a}$ without invoking the lapses. Only in the case where $c_1 c_2 c_3\ne 0$ and when $\mathcal{A}\ne 0$, the rotations depend non--trivially on the lapses.  Integrating out the rotations we find that the Hamiltonian is no longer linear in the lapses and the theory excites the BD ghost
\ba
&&H=N_1 R_{1,0}+N_2 R_{2,0}+2m^2\mpl \Big\{ c_3 \sqrt{g_2}+c_1[E_1]\nn\\
&+&N_1(c_1+c_2  \sqrt{G_{12}})+N_2(c_2\sqrt{g_1}+c_3[E_2])\nn\\[3pt]
&+&\frac{c_2^2N_1^2\(c_1[E_1]+c_3 N_2[E_2]\)\mathcal A^2}{2\(c_1 c_3 N_2[E_1][E_2]+c_2 \sqrt{G_{12}}N_1\(c_1[E_1]+c_3 N_2[E_2]\)\)}\nn\\[3pt]
&+&\mathcal{O}\(\mathcal{A}^3\)\Big\} +\mathcal{O}\(R_i\)\,.
\ea
The contributions arising from integrating out the boosts will all enter at least at first order in the momentum constraints $R_i$ and cannot change the nature of the result obtained so far. We can see straight away that if we break the symmetric vielbein condition and $\mathcal{A}\ne 0$, then as soon as $c_1 c_2 c_3\ne 0$, the second line leads to terms which are highly  non--linear in the lapses and thus spoil the constraints that should have been there to remove the BD ghost. The only possible way out is then to solve the equations of motion by imposing $\mathcal{A}=E^{\ \ a}_{1,[1}E^{\ \ b}_{1,2]}\delta_{ab}=0$. However this would overconstrain the system, and does not follow from any combination of equations of motion.
As a result we confirm that cycles of interactions in multi--gravity break the symmetric vielbein condition and also lead to a BD ghost  in the unconstrained vielbein language. This is consistent with the conclusions of \cite{Afshar:2014dta}.

\para{Discussion}
To summarize, the composite vielbein coupling is a ghost--free matter coupling in the decoupling limit, where it is equivalent to the composite metric coupling considered in \cite{deRham:2014naa}, however beyond the decoupling limit it propagates a BD ghost, independently on whether we consider it in the metric language or in the unconstrained vielbein one. Although we have not considered more general composite vielbein couplings here, any other composite metric or vielbein will lead to a ghost already in the decoupling limit\footnote{This can be shown by going to Jordan frame in the vielbein language. If the transformation to Jordan frame involves anything other than a linear vielbein redefinition $e^a \to e^a - \alpha f^a$, the mass term will no longer be of the acceptable form and will lead to a ghost already at the scale $\Lambda$ \cite{deRham:2010ik}.}.\\

 These results may viewed positively in two ways, on the one hand there do exist ghost--free matter couplings in the decoupling limit, and that is a sufficient for validity of the EFT for energies $E \le \Lambda$. On the other hand the absence of any non--minimal matter coupling, like the absence of non--minimal kinetic terms \cite{deRham:2013tfa}, is a remarkable testament to the power of field theory consistency in determining unique interacting theories.\\

This analysis has also allowed us to study more general theories where the symmetric vielbein condition is broken and the vielbein formulation differs from the metric one. We argue that in that case the BD is also present in the vielbein formulation of the theory. Taken together with the results of \cite{deRham:2015rxa} we are led to conclude that any consistent interacting theory for massive spin-two fields must have a natural expression in the metric language and that the vielbein formalism, although a very powerful tool from the calculational point of view and a very convenient formalism, does not give rise to new interactions.\\

{\bf Acknowledgments:}
We would like to thank Kurt Hinterbichler, Andrew Matas, Nick Ondo, Rachel Rosen and Shuang--Yong Zhou for useful discussions. CdR is supported by a Department of Energy grant DE--SC0009946. AJT is supported by Department of Energy Early Career Award DE--SC0010600.

\bibliography{references}

\end{document}